\documentclass[aps,twocolumn,superscriptaddress]{revtex4-2}

\usepackage{graphicx}  
\usepackage{float}     
\usepackage{dcolumn}   
\usepackage{bm}        
\usepackage{amssymb}   
\usepackage{amsmath}   
\usepackage{lipsum}    
\usepackage{xcolor}
\usepackage{soul}
\usepackage{textcomp}
\usepackage{cancel}

\usepackage[utf8]{inputenc}
\usepackage{natbib}
\usepackage{graphicx}

\definecolor{goodred}{RGB}{183,15,58}
\definecolor{goodblue}{RGB}{93,128,180}
\definecolor{goodgreen}{RGB}{95,118,33}

\usepackage{color}

\begin{document}
\title{Toward spin-inertial magnonics with ferromagnets}

\author{Mikhail Cherkasskii}
\email[]{macherkasskii@hotmail.com}
\affiliation{Faculty of Physics, University of Duisburg-Essen, Duisburg, 47057, Germany}

\author{Igor Barsukov}
\email[]{igorb@ucr.edu}
\affiliation{Physics and Astronomy, University of California, Riverside, California 92521, United States}

\author{Andrey Stashkevich}
\affiliation{Universit\'e Sorbonne Paris Nord, LSPM, CNRS, UPR 3407, F-93430, Villetaneuse, France}

\author{Eloi Haltz}
\affiliation{Universit\'e Sorbonne Paris Nord, LSPM, CNRS, UPR 3407, F-93430, Villetaneuse, France}

\author{Yves Roussign\'e}
\affiliation{Universit\'e Sorbonne Paris Nord, LSPM, CNRS, UPR 3407, F-93430, Villetaneuse, France}

\author{Rundong Jiang}
\affiliation{Physics and Astronomy, University of California, Riverside, California 92521, United States}

\author{Svetlana Sheshukova}
\affiliation{Saratov State University, 410012 Saratov, Russia}

\author{Alexandr V. Sadovnikov}
\email[]{sadovnikovav@gmail.com}
\affiliation{Saratov State University, 410012 Saratov, Russia}

\begin{abstract}
Ferromagnetic magnonics is conventionally limited to the gigahertz range of precessional spin dynamics. Spin inertia offers a route beyond this constraint by introducing an additional high-frequency branch of magnetization dynamics. So far, this inertial mode has been observed only in the spatially uniform limit, leaving open whether it can propagate. Introducing wave-vector-resolved Brillouin light scattering as a probe of inertial dynamics, we demonstrate propagating inertial spin waves in a ferromagnetic thin film. For propagation perpendicular to the magnetization, they form chiral surface waves with field-reversible nonreciprocity. Their group velocity exceeds exchange-based expectations by about two orders of magnitude, pointing to a nonlocal contribution to spin inertia. These findings establish spin-inertial magnonics as an experimentally accessible route to terahertz wave functionality, combining ultrahigh frequencies with the mature materials, interfaces, and device engineering of ferromagnets.
\end{abstract}
\maketitle
By exploiting phase coherence, interference, nonreciprocity, and tunable dispersion, magnonic systems enable wave-based 
functionalities for logic operations and signal routing that hold promise for conventional, neuromorphic, and quantum information technologies \cite{Chumak22}.

The most extensive experimental progress in magnonics has been achieved with ferromagnets, where refined techniques for spin-wave excitation and manipulation have been established over decades \cite{nikolaev26,Chumak2015}. These magnonic realizations rely on spin waves governed by precessional magnetization dynamics. Thus, pushing these systems to higher frequencies and faster operation speeds requires very large magnetic fields incompatible with device architectures, or very short wavelengths that impose severe challenges on device fabrication and operation.

A particularly appealing route beyond these constraints is offered by a phenomenon in which a spin system acquires inertia due to coupling to a bath of other degrees of freedom~\cite{quarenta2024_bathinduced, bhattacharjee2012_atomistic, mondal2017_relativistic,kikuchi2015_spin,fahnle2011_generalized,NikolicPRL26,Moussa}. This spin inertia enters the equation of magnetization motion as a term proportional to the second-order time derivative of the order parameter, reflecting the finite response time of angular momentum transfer between spins and their environment \cite{de2025_magnetic}. In contrast to earlier notions of magnetic inertia associated with the effective mass of translating or deforming spin textures \cite{Makhfudz12,buettner15}, spin inertia endows even a spatially uniform ferromagnetic order with an additional dynamical degree of freedom.

The striking consequence of spin inertia is the emergence of a collective excitation at frequencies far above the precessional gigahertz range. This inertial mode lifts the aforementioned requirement for prohibitively large magnetic fields or short wavelengths. It opens a qualitatively new dynamical regime at THz frequencies \cite{rodriguez2024_spin}, while benefiting from the established techniques of material, device, and spin-torque engineering associated with ferromagnets \cite{nikolaev26,Chumak2015,gonccalves2018oscillatory,BarsukovMagneticPhase,Zhang2010,MeckecstockAPL}.

Despite recent experimental progress, all observations of spin-inertial dynamics have so far been restricted to spatially uniform excitations~\cite{neeraj2021_inertial,unikandanunni2022_inertial,de2025_magnetic}. These experimental studies have established the existence of the high-frequency inertial mode but left open the question of whether spin-inertial dynamics can support propagating spin waves at finite wave vectors.

Finite-wave-vector excitations are essential for magnonics, as propagation, dispersion engineering, and nonreciprocity require collective modes that carry momentum. Demonstrating propagating inertial spin waves is therefore a critical step toward establishing spin-inertial dynamics as a platform for wave-based information processing.

\begin{figure*}[!hbtp]
    \centering
    \includegraphics[width=\textwidth]{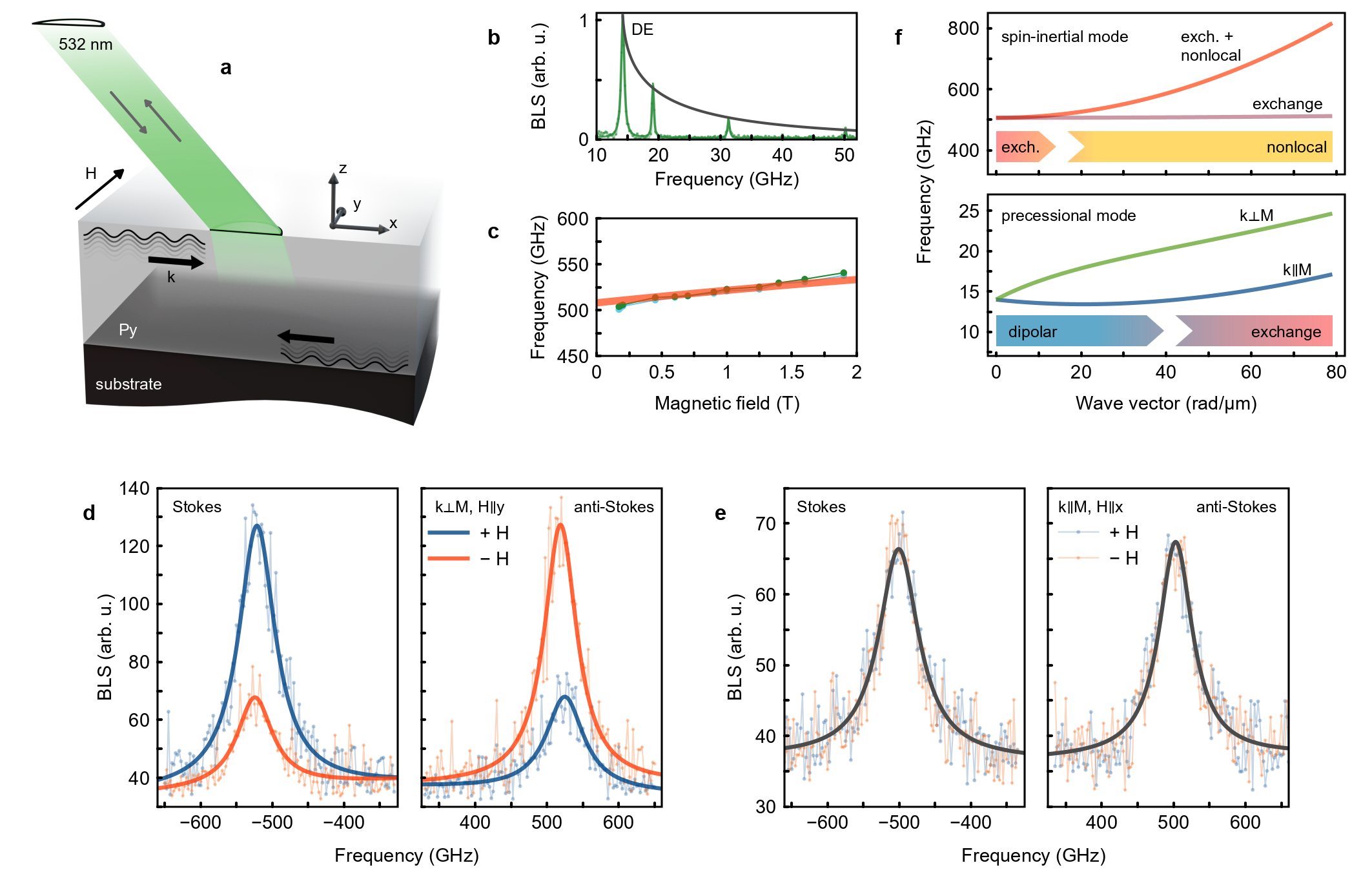}
    \caption{
(a)~BLS setup in the backscattering geometry. (b)~Low-frequency spectrum of precessional standing spin waves. The intensity decays rapidly with increasing frequency (mode number). (c)~Frequency vs. field relation of the inertial mode with the secular-equation fit. (d)~BLS spectrum of the inertial mode propagating perpendicular to the equilibrium magnetization. The Stokes/anti-Stokes intensity asymmetry reverses upon reversing the field polarity. (e)~In a control experiment with the magnetic field along the $x$-axis, the inertial mode propagating along the equilibrium magnetization shows no Stokes/anti-Stokes asymmetry. (f)~Schematic of the two spin-wave branches. The lower-frequency precessional spin waves are dominated by dipolar interaction at low wave vectors and by exchange at high wave vectors. The higher-frequency inertial spin waves are dominated by spatial nonlocality already at relatively small wave vectors.}
    \label{fig:NutSchematic}
\end{figure*}

Here, we establish Brillouin light scattering for accessing spin-inertial dynamics at finite wave vectors in a ferromagnetic thin film. We directly measure inertial spin waves and establish their propagating nature, including signatures of nonreciprocal transport. Resolving their dispersion relation reveals an anomalously enhanced group velocity, which points to a nonlocal contribution to spin inertia. Our results establish spin-inertial magnonics as a promising dynamical regime and open a route toward magnonic operation at \mbox{(sub-)}terahertz frequencies with ferromagnets.

\textbf{Brillouin scattering by the inertial mode}

In Brillouin light scattering (BLS), an incident photon is inelastically scattered by creating or annihilating a spin-wave quantum (magnon) \cite{nikolaev26,DEMOKRITOV2001441,Carlotti2002BLS}. In conventional implementations, this process is used to probe gigahertz-frequency precessional spin waves. Here, we enhance the BLS spectroscopy with a tandem Fabry--P\'erot interferometer \cite{Scarponi}, providing sufficient spectral range for the higher-frequency inertial mode.

We employ backscattering geometry, where scattered light is collected along the axis of the incident laser beam, as shown in Fig.\,1a. By varying the incidence angle, we can continuously tune the momentum transferred to the spin system, allowing us to resolve both the frequency $f$ and the wave vector $k$ of magnons propagating along the x-axis. We probe the room-temperature spectra of a ferromagnetic thin film (Permalloy, Py). Its magnetization $\bf{M}$ is saturated along the $y$-axis, within the film plane, by a magnetic field $H$. 

The low-frequency region of the BLS spectrum shows, as expected, a precessional mode. In the experimental configuration of Fig.\,1a with $\mathbf{k}\perp\mathbf{M}$, it can be identified as the well-known Damon-Eshbach (DE) mode. 
With increasing frequency, standing waves \cite{KalinikosSlavin1986} form across the film thickness. Their BLS intensities decrease rapidly with increasing out-of-plane mode number, rendering higher-order modes experimentally undetectable (Fig.\,1b).

However, when we increase the BLS probing frequency further still, we find a pronounced peak around 0.5\,THz, shown in Fig.\,1d. Its intensity is several orders of magnitude larger than that of the precessional spin waves at the floor of the BLS spectrum. It cannot be attributed to precessional dynamics.

The peak frequency lies close to that of the inertial mode previously observed in Permalloy using time-domain measurements with frequency-selective terahertz-field excitation \cite{neeraj2021_inertial} -- a fundamentally different excitation and detection technique. 
First, we restrict our measurements to a small wave vector. With increasing magnetic field, the peak systematically blue-shifts. To our knowledge, this is the first direct measurement of the magnetic-field dependence of the inertial mode. As shown in Fig.\,1c, the observed blue shift agrees with the secular equation of inertial spin dynamics \cite{cherkasskii2022_theory,rodriguez2024_spin}.

The magnetic-field dependence of the peak frequency already rules out nonmagnetic excitations such as acoustic waves; nevertheless, we briefly consider possible phononic contributions for completeness. Propagating acoustic phonons have a prohibitively large wave-vector mismatch for BLS at high frequencies. They can form standing-wave modes across the film thickness or the thickness of the substrate. Such confined acoustic excitations generally form thickness-dependent sequences of resonances, rather than a single isolated mode \cite{Zhang2003}. In our BLS setup, phononic contributions are suppressed by polarization selection \cite{Demokritov1994} -- no indication of a phonon sequence is observed in the spectra.

We further perform an experiment on a control sample, with a different film thickness and substrate. These changes substantially modify the phononic mode spectrum, yet the same peak is observed. Together with its magnetic-field dependence and frequency scale, these results strongly support the identification of the BLS peak with the inertial mode and motivate the next question: whether this mode can propagate.

\textbf{Nonreciprocal inertial spin waves}

When the wave vector transferred in the BLS process is increased, the inertial peak remains observable in the spectra, which shows that the inertial mode in fact carries momentum. Although typically visualized as nutational motion superimposed on top of precessional motion, spin-inertial dynamics is not merely a correction to precession. It should rather be viewed as an independent, higher-frequency branch of spin waves, as illustrated in Fig.\,1f.

We find that the Stokes peak is substantially larger than the anti-Stokes peak, as shown in Fig.\,1d. Upon reversal of the magnetic field, this intensity asymmetry reverses -- the anti-Stokes peak becomes larger than the Stokes peak.

To explain this observation, we start with the inertial equation \cite{ciornei2011_magnetization} of magnetization motion \cite{wegrowe2012_magnetization,cherkasskii2022_theory,rodriguez2024_spin},
\begin{equation} \label{eq:ILLG}
\begin{split}
\partial_t \mathbf{m} =
- |\gamma| \mu_0 \mathbf{m} \times \mathbf{H}_{\mathrm{eff}}
+ \alpha \mathbf{m} \times \partial_t \mathbf{m}
+ \eta \mathbf{m} \times \partial_{tt} \mathbf{m},
\end{split}
\end{equation}
where $\mathbf{m}=\mathbf{M}/M$, ${\bf{H}}_{{\rm{eff}}}$ is the effective field, $\alpha$ is the Gilbert damping, $\eta$ is the inertial parameter, and $\gamma$ is the gyromagnetic ratio. We implement dynamic dipole interaction by extending Kalinikos' formalism of tensor magnetostatic Green's functions \cite{KalinikosSlavin1986}.

For the experimental geometry shown in Fig.\,1a, this theoretical framework results in an inertial, high-frequency analog of the Damon-Eshbach spin waves \cite{DamonEshbach1961,Hertel,Titov}
. By virtue of the dipole field, inertial spin waves propagating along the $x$-axis are localized at the top and bottom surfaces of the thin film. They obey the relationship
\begin{equation} \label{right-hand}
\mathbf{k}\uparrow\!\uparrow \mathbf{m} \times \mathbf{n}, 
\end{equation}
where $\mathbf{n}$ is the outward normal vector of the surface at which the spin wave is localized.

The observed intensity asymmetry of Stokes and anti-Stokes peaks directly corroborates this feature of inertial spin waves. The asymmetry arises from the unequal overlap of the probing laser field with the surface-localized spin waves, as shown in Fig.\,1a. The optical field penetrates only about 10\,nm deep into the film (exponential intensity attenuation), resulting in a stronger coupling to spin waves localized at the top surface than at the bottom surface \cite{Camley1982}.

For a positive field $H$, the Stokes process at the top surface is dominant while the anti-Stokes process at the bottom surface is partially suppressed due to the reduced overlap. Upon reversal of the magnetic field, magnetization $\bf{m}$ and consequently the wave vector $\bf{k}$ in Eq.\,\ref{right-hand} flip sign, while the optical probing conditions remain unchanged. The intensity asymmetry of the Stokes and anti-Stokes peaks inverts. With increasing wave vector, the surface localization is expected to become more pronounced. Consistent with this expectation \cite{Stashkevich2007}, we observe an increasing intensity asymmetry between the Stokes and anti-Stokes peaks.

In a control experiment, we align the magnetic field along the $x$-axis. Now, the sample is magnetized in the direction of the wave vector probed by BLS, $\mathbf{k}\parallel\mathbf{M}$. Our theoretical framework anticipates that such inertial spin waves are not surface-localized. As shown in Fig.\,1e, the BLS spectra confirm no asymmetry of Stokes and anti-Stokes intensities for both polarities of the magnetic field.

These results establish the propagating nature of inertial spin waves and lay the foundation for ultrahigh-frequency spin-inertial magnonics. Of particular interest are the inertial surface-localized spin waves propagating perpendicular to the magnetization, $\mathbf{k}\perp\mathbf{M}$. The relation in Eq.\,\ref{right-hand} reflects their chiral character: for a given magnetization direction, spin waves with opposite wave vectors are propagating at opposite film surfaces. This nonreciprocity enables directional transport of inertial spin waves, a property that is sought after for advanced magnonic applications \cite{Chumak22}. For further functionalization of these inertial spin waves, we must yet establish their dispersion relation.

\textbf{Enhanced group velocity}

Figure~2a shows the frequency of the inertial spin waves as a function of the wave vector; it follows a parabola. We find that the dispersion relations of the spin waves counter-propagating at the top and bottom surfaces of the film, corresponding to the Stokes and anti-Stokes peaks, follow nearly the same behavior, i.e. $f(k)\approx f(-k)$.

\begin{figure}[t]
    \centering
    \includegraphics[width=0.45\textwidth]{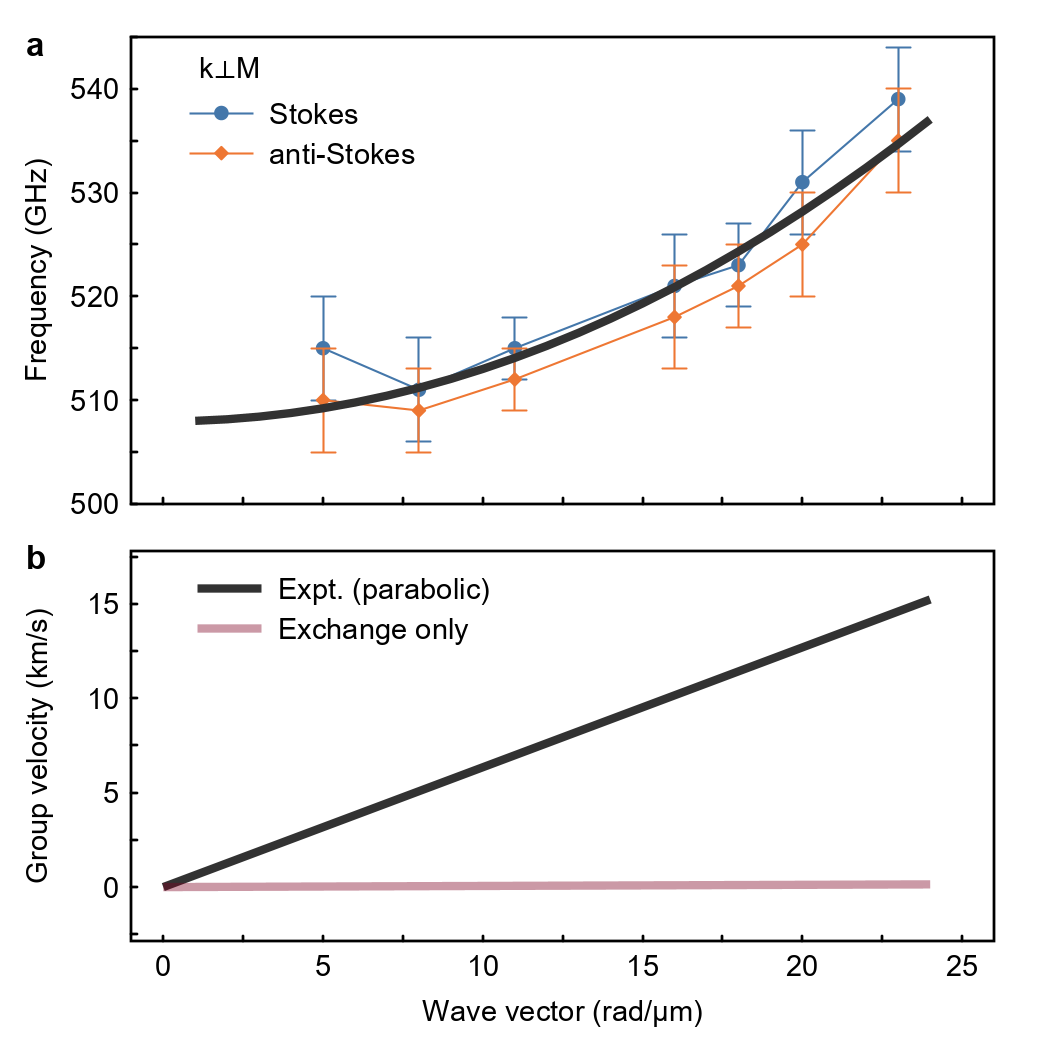}
    \caption{
(a)~Frequency of the inertial mode from Stokes and anti-Stokes peaks, as a function of the wave vector. The gray curve is a parabolic fit. (b)~Group velocity of Damon-Eshbach inertial spin waves, inferred from the parabolic fit, is much larger than the group velocity expected from exchange interaction in the absence of spatial nonlocality.}
    \label{fig:2}
\end{figure}

Based on the theoretical framework, the dispersion relation of inertial spin waves is $2\pi\,f=\eta^{-1}+ \mathcal{O}(H)+D_{\omega}k^2$. While the dipolar interaction localizes the spin waves at the film surfaces in the Damon-Eshbach configuration, it has a minuscule effect on their frequency. The group velocity is thus expected to increase linearly with the wave vector as $v_\mathrm{g}=2\pi\,\partial f/\partial k=2D_{\omega}k$, i.e. the same dependence as for precessional spin waves in the exchange regime. We determined the spin-wave stiffness $D_{\omega}\approx \gamma\cdot(2\cdot10^{-17}$\,T\,m$^2)$ of our Permalloy samples from measurements of the precessional standing spin waves performed at the beginning of this study.

However, when evaluating the group velocity of the inertial spin waves in Fig.\,2b, we find a markedly different behavior. The slope of the group velocity is about 100 times larger than expected from the measured spin-wave stiffness. We observe a similarly large enhancement of the group velocity in the previously mentioned control sample as well. This result is significant from a technological standpoint. We find that inertial spin waves in ferromagnets combine ultrahigh frequencies with group velocities comparable to those of antiferromagnetic magnons \cite{KimmelNewton,Hamdi23,Haiming23}.

\textbf{Nonlocal contributions}

Current theoretical descriptions \cite{MondalLevente,cherkasskii2024_inertial,Serpico,Lomonosov,Yuan} of spin inertia do not, by themselves, account for the enhanced group velocity observed here, but they suggest a natural direction for its explanation. Rather than being an independent phenomenological modification of magnetization dynamics, spin inertia can be viewed as part of a broader description of magnetic dynamics with memory \cite{NikolicPRL26}. In retarded-damping formulations, the instantaneous, or Markovian, limit yields the usual Gilbert torque \cite{hartmann2025_intrinsic}, while the next term in the low-frequency expansion of the damping kernel gives the inertial contribution proportional \cite{Anders2025} to ${\partial _{tt}} \bf{m}$. 

If spin inertia emerges from a retarded spin–environment response, then it should not  be treated as a purely local material parameter. Indeed, when magnetic damping is derived from the coupling of the magnetization to electronic or other bath degrees of freedom, the resulting damping torque is generally nonlocal in both time and space \cite{NikolicPRL26}. The temporal nonlocality gives rise, in a low-frequency expansion, to the conventional Gilbert term (and to the inertial correction). The spatial nonlocality, in turn, leads to gradient corrections to the damping torque and hence to wave-vector-dependent damping.

For the Gilbert damping parameter, this idea is well established. Theoretical work \cite{Tserkovnyak2009} has shown that nonuniform magnetization dynamics can generate gradient corrections to the damping torque, including terms of the form $\mathbf{m}\times\nabla^2\partial_t{\mathbf{m}}$, which for long-wavelength spin waves correspond to an expansion $\alpha(k)=\alpha_0+\alpha_2 k^2+\mathcal{O}(k^4)$. Experimentally, mode- and wave-vector-dependent damping has been observed in nanomagnets and perpendicular standing spin-wave resonances \cite{Nembach2013,Li_2016}.

Spatially nonlocal magnetic inertia has been derived microscopically from the coupling of localized spins to itinerant electrons \cite{NikolicPRL26}. 
In the long-wavelength limit, such nonlocality can be represented by a wave-vector-dependent inertial parameter, $\eta \rightarrow \eta(k)=\eta_0+\eta_2 k^2+\mathcal{O}(k^4)$. This correction modifies the curvature of the inertial dispersion and, consequently, the group velocity
\begin{equation}    \label{v_g}
    v_\mathrm{g} \approx 2 D_{\omega}k - \frac{2\eta_2}{\eta_0^2} k.
\end{equation}

Spatially nonlocal Gilbert damping \cite{NikolicPRL26,Li_2016} provides a useful precedent for wave-vector-dependent dissipative torques, although its gradient correction \cite{Hankiewicz2008,Tserkovnyak2009} is material- and mechanism-dependent and need not have the same sign as the inertial correction. The extracted inertial correction is negative, $\eta_2<0$, indicating that the effective spin inertia decreases with increasing wave vector (and enhances the group velocity). This behavior is consistent with a spatially extended inertial response whose Fourier amplitude decreases away from the uniform limit. The strength of this nonlocal contribution can be characterized by a length scale,
\begin{equation}
    \ell_\eta = \sqrt{\left|\eta_2/\eta_0\right|} \approx 10\text{\textendash}20\,\mathrm{nm}.
\end{equation}
Within the experimentally accessible wave-vector range, this corresponds to a modest variation of the inertial parameter. Nevertheless, owing to the high characteristic frequency of the inertial mode, this variation produces a pronounced enhancement of its dispersion curvature and group velocity observed in the experiment.

\textbf{Conclusions}

In summary, we report the first observation of spin-inertial dynamics by Brillouin light scattering, establishing BLS as a versatile probe of this regime. 
Crucially, the intrinsic wave-vector selectivity of BLS allows us to move beyond the uniform limit and demonstrate that the inertial mode forms a genuine propagating spin-wave branch carrying momentum. For propagation perpendicular to the magnetization, the field-reversible Stokes/anti-Stokes asymmetry reveals its chiral surface-wave character, with counter-propagating inertial spin waves localized at opposite film interfaces.

The measured dispersion reveals an additional and unexpected advantage of this spin-wave branch. Its curvature cannot be accounted for by the exchange stiffness alone: the observed group velocity exceeds the exchange-based expectation by approximately two orders of magnitude. Within a long-wavelength description, this enhancement can be captured by a nonlocal contribution to spin inertia. Our results point to spin inertia as part of a retarded and spatially nonlocal coupling between the magnetization and its environment. The experimentally resolved dispersion provides a benchmark for microscopic descriptions of high-frequency spin dynamics that retain the full spatial and temporal nonlocality of the spin–environment response \cite{NikolicPRL26,weissenhofer2026_spurious,hartmann2025_intrinsic,beater2026}, including the possibility that high-frequency spectral features reflect bath memory beyond the inertial truncation \cite{weissenhofer2026_spurious}.

Our findings establish spin-inertial magnonics as an experimentally accessible route to field-tunable ultrahigh-frequency wave functionality. This route combines terahertz dynamics with the mature toolbox of materials, interfaces, spin-torque excitation, and device engineering available for ferromagnets. The chiral surface localization demonstrated here may be exploited for directional signal routing, isolation, and other nonreciprocal wave-based functionalities. The high group velocity shortens signal transit times, while the steep dispersion enhances the available bandwidth for signal processing at the terahertz frequencies targeted by future wireless communication. Future work should establish the microscopic origin of the nonlocal inertial response and determine how it depends on material properties and sample morphology.

\section*{Methods}
\textbf{Samples}
The sample was grown by sputtering 50\,nm polycrystalline Permalloy (Py, Ni$_{81}$Fe$_{19}$) and a 2\,nm Ta capping layer onto a Si substrate. The control sample was a 20\,nm single-crystal Py with a 1.7\,nm Al capping layer grown on an MgO substrate.

\textbf{BLS measurements}
The BLS measurements were carried out at room temperature, using a single-mode 532\,nm laser. The vertically polarized beam was passed through a temperature-controlled etalon, suppressing laser sidelobes by a factor of 600, and focused onto the sample to a spot size of $40\,\mu$m. The backscattered light was analyzed by a TFP-2 HC tandem Fabry--P\'erot interferometer \cite{Scarponi}, a high-contrast variant of the Sandercock-type multipass interferometer \cite{NizzoliSandercock1990}. Polarization and retardation optics at each pass of the advanced $3\times3$-pass configuration decouple successive passes, while the tandem arrangement suppresses side-order replicas and extends the accessible spectral range. The mirror spacing for the high-frequency measurements was set to 0.20\,mm~\cite{Demokritov1994}. The high contrast of the TFP-2 HC suppresses the elastic and air-scattering background; polarization selection additionally rejects phonon contributions \cite{NizzoliSandercock1990}.

\section*{Acknowledgments}
The authors thank Anna Semisalova and Michael Farle for fruitful discussions, Alexandr P. Nosov for providing the samples, and Noah Townsend-Rodnunsky for assistance in preparing the figures.

\section*{Author contributions}
M.C. organized and coordinated this collaboration. S.S. and A.V.S. carried out the experimental studies. A.S., E.H., and Y.R. provided expertise in BLS and spin-wave modeling. R.J. assisted in calculating the nutational response. I.B. provided overall scientific leadership and led the writing of the manuscript, to which all authors contributed.

\section*{Funding}
I.B.'s research was supported by funds from the UC National Laboratory Fees Research Program of the University of California, Grant Number L25CR9003. 

\section*{Competing interests}
The authors declare no competing interests.
\section*{Data availability}
The data that support the findings of this study are available from the corresponding authors upon reasonable request.

\bibliography{references}

@PREAMBLE{
 "\providecommand{\noopsort}[1]{}" 
 # "\providecommand{\singleletter}[1]{#1}%" 
}

@article{nikolaev26,
  author  = {Nikolaev, K. O. and Raskhodchikov, D. and Bensmann, J. and Niehues, S. and Bessonov, V. D. and Lomonte, E. and Jin, L. and Schmidt, R. and Kern, J. and {Michaelis de Vasconcellos}, S. and Bratschitsch, R. and Demokritov, S. O. and Pernice, W. H. P. and Demidov, V. E.},
  title   = {Zero-field spin-wave networks with integrated non-volatile memory},
  journal = {Nat. Mater.},
  year    = {2026},
  doi = {10.1038/s41563-026-02778-1},
}

@incollection{NizzoliSandercock1990,
  author    = {Nizzoli, F. and Sandercock, J. R.},
  booktitle = {Dynamical Properties of Solids},
  editor    = {Horton, G. K. and Maradudin, A. A.},
  volume    = {6},
  pages     = {281--335},
  publisher = {North-Holland},
  address   = {Amsterdam},
  year      = {1990}
}

@article{Demokritov1994,
  title = {Light scattering from spin waves in thin films and layered systems},
  volume = {6},
  ISSN = {1361-648X},
  url = {http://dx.doi.org/10.1088/0953-8984/6/36/002},
  DOI = {10.1088/0953-8984/6/36/002},
  number = {36},
  journal = {J. Phys.: Condens. Matter},
  publisher = {IOP Publishing},
  author = {Demokritov,  S and Tsymbal,  E},
  year = {1994},
  month = Sept,
  pages = {7145–7188}
}

@article{Anders2025,
  author  = {Hartmann, Felix and Tietjen, Finja and Geilhufe, R. Matthias and Anders, Janet},
  title   = {Quantifying non-{Markovianity} in magnetization dynamics via entropy production rates},
  journal = {Commun. Phys.},
  volume  = {9},
  pages   = {292},
  year    = {2026}
}

@article{buettner15,
  author = {B{\"u}ttner, Felix and Moutafis, C.
            and Schneider, M. and Kr{\"u}ger, B.
            and G{\"u}nther, C. M. and Geilhufe, J.
            and von Korff Schmising, C. and Mohanty, J.
            and Pfau, B. and Schaffert, S.
            and Bisig, A. and Foerster, M.
            and Schulz, T. and Vaz, C. A. F.
            and Franken, J. H. and Swagten, H. J. M.
            and Kl{\"a}ui, M. and Eisebitt, S.},
  title = {Dynamics and inertia of skyrmionic spin structures},
  journal = {Nat. Phys.},
  volume = {11},
  pages = {225--228},
  year = {2015},
  doi = {10.1038/nphys3234}
}

@article{MondalLevente,
  title = {Inertial spin waves in ferromagnets and antiferromagnets},
  author = {Mondal, Ritwik and R{\'o}zsa, Levente},
  journal = {Phys. Rev. B},
  volume = {106},
  issue = {13},
  pages = {134422},
  numpages = {14},
  year = {2022},
  month = {Oct},
  publisher = {American Physical Society},
  doi = {10.1103/PhysRevB.106.134422},
  url = {https://link.aps.org/doi/10.1103/PhysRevB.106.134422}
}

@article{Serpico,
  title = {Micromagnetic study of inertial spin waves in ferromagnetic nanodots},
  author = {d'Aquino, Massimiliano and Perna, Salvatore and Pancaldi, Matteo and Hertel, Riccardo and Bonetti, Stefano and Serpico, Claudio},
  journal = {Phys. Rev. B},
  volume = {107},
  issue = {14},
  pages = {144412},
  numpages = {8},
  year = {2023},
  month = {Apr},
  publisher = {American Physical Society},
  doi = {10.1103/PhysRevB.107.144412},
  url = {https://link.aps.org/doi/10.1103/PhysRevB.107.144412}
}

@article{Hertel,
  title = {Nonreciprocal Inertial Spin-Wave Dynamics in Twisted Magnetic Nanostrips},
  author = {d'Aquino, Massimiliano and Hertel, Riccardo},
  journal = {Phys. Rev. Lett.},
  volume = {135},
  issue = {21},
  pages = {216705},
  numpages = {7},
  year = {2025},
  month = {Nov},
  publisher = {American Physical Society},
  doi = {10.1103/gvwf-fsz6},
  url = {https://link.aps.org/doi/10.1103/gvwf-fsz6}
}

@article{Lomonosov,
  title = {Anatomy of inertial magnons in ferromagnetic nanostructures},
  author = {Lomonosov, Alexey M. and Temnov, Vasily V. and Wegrowe, Jean-Eric},
  journal = {Phys. Rev. B},
  volume = {104},
  issue = {5},
  pages = {054425},
  numpages = {7},
  year = {2021},
  month = {Aug},
  publisher = {American Physical Society},
  doi = {10.1103/PhysRevB.104.054425},
  url = {https://link.aps.org/doi/10.1103/PhysRevB.104.054425}
}

@article{Titov,
  title = {Nutation spin waves in ferromagnets},
  author = {Titov, Sergei V. and Dowling, William J. and Kalmykov, Yuri P. and Cherkasskii, Mikhail},
  journal = {Phys. Rev. B},
  volume = {105},
  issue = {21},
  pages = {214414},
  numpages = {9},
  year = {2022},
  month = {Jun},
  publisher = {American Physical Society},
  doi = {10.1103/PhysRevB.105.214414},
  url = {https://link.aps.org/doi/10.1103/PhysRevB.105.214414}
}

@article{Moussa,
  title = {Unquenched orbital angular momentum as the origin of spin inertia},
  author = {Moussa, Tarek and Basu, Darpa Narayan and Mondal, Ritwik and Kamra, Akashdeep},
  journal = {Phys. Rev. B},
  volume = {114},
  issue = {2},
  pages = {024424},
  numpages = {16},
  year = {2026},
  month = {Jul},
  publisher = {American Physical Society},
  doi = {10.1103/9h9p-8qzd},
  url = {https://link.aps.org/doi/10.1103/9h9p-8qzd}
}

@article{Yuan,
  title = {Using surface plasmons to detect spin inertia},
  author = {Yuan, H. Y.},
  journal = {Phys. Rev. B},
  volume = {112},
  issue = {5},
  pages = {054438},
  numpages = {7},
  year = {2025},
  month = {Aug},
  publisher = {American Physical Society},
  doi = {10.1103/5dgr-4tkt},
  url = {https://link.aps.org/doi/10.1103/5dgr-4tkt}
}

@article{Zhang2010,
  title = {Magnetocrystalline anisotropy of {Fe-Si} alloys on {MgO(001)}},
  author = {Zhang, Y. N. and Cao, J. X. and Barsukov, I. and Lindner, J. and Krumme, B. and Wende, H. and Wu, R. Q.},
  journal = {Phys. Rev. B},
  volume = {81},
  issue = {14},
  pages = {144418},
  numpages = {6},
  year = {2010},
  month = {Apr},
  publisher = {American Physical Society},
  doi = {10.1103/PhysRevB.81.144418},
  url = {https://link.aps.org/doi/10.1103/PhysRevB.81.144418}
}

@article{MeckecstockAPL,
    author = {Meckenstock, Ralf and Barsukov, I. and Posth, O. and Lindner, J. and Butko, A. and Spoddig, D.},
    title = {Locally resolved ferromagnetic resonance in {Co} stripes},
    journal = {Appl. Phys. Lett.},
    volume = {91},
    number = {14},
    pages = {142507},
    year = {2007},
    month = {10},
    issn = {0003-6951},
    doi = {10.1063/1.2794026},
    url = {https://doi.org/10.1063/1.2794026},
}

@article{DEMOKRITOV2001441,
title = {Brillouin light scattering studies of confined spin waves: linear and nonlinear confinement},
journal = {Phys. Rep.},
volume = {348},
number = {6},
pages = {441-489},
year = {2001},
issn = {0370-1573},
doi = {https://doi.org/10.1016/S0370-1573(00)00116-2},
url = {https://www.sciencedirect.com/science/article/pii/S0370157300001162},
author = {S.O. Demokritov and B. Hillebrands and A.N. Slavin}
}

@article{Carlotti2002BLS,
  author  = {Carlotti, G. and Gubbiotti, G.},
  title   = {Magnetic properties of layered nanostructures studied by means of Brillouin light scattering and the surface magneto-optical Kerr effect},
  journal = {J. Phys.: Condens. Matter},
  volume  = {14},
  pages   = {8199--8233},
  year    = {2002},
  doi     = {10.1088/0953-8984/14/35/303}
}

@article{Stashkevich2007,
  author  = {Stashkevich, A. A. and Djemia, P. and
             Fetisov, Y. K. and Bizi{\`e}re, N. and
             Fermon, C.},
  title   = {High-intensity Brillouin light scattering by
             spin waves in a permalloy film under
             microwave resonance pumping},
  journal = {J. Appl. Phys.},
  volume  = {102},
  pages   = {103905},
  year    = {2007},
  doi     = {10.1063/1.2815673}
}

@article{Haiming23,
  title = {Long-Distance Coherent Propagation of High-Velocity Antiferromagnetic Spin Waves},
  author = {Wang, Hanchen and Yuan, Rundong and Zhou, Yongjian and Zhang, Yuelin and Chen, Jilei and Liu, Song and Jia, Hao and Yu, Dapeng and Ansermet, Jean-Philippe and Song, Cheng and Yu, Haiming},
  journal = {Phys. Rev. Lett.},
  volume = {130},
  issue = {9},
  pages = {096701},
  numpages = {7},
  year = {2023},
  month = {Feb},
  publisher = {American Physical Society},
  doi = {10.1103/PhysRevLett.130.096701},
  url = {https://link.aps.org/doi/10.1103/PhysRevLett.130.096701}
}

@article{KimmelNewton,
title = {Nanoworld of terahertz magnetism opening new frontiers},
journal = {Newton},
volume = {2},
number = {9},
pages = {100521},
year = {2026},
issn = {2950-6360},
doi = {https://doi.org/10.1016/j.newton.2026.100521},
url = {https://www.sciencedirect.com/science/article/pii/S2950636026001234},
author = {Alexey V. Kimel and Boris A. Ivanov and Dmytro Afanasiev}
}

@article{Hankiewicz2008,
  title = {Inhomogeneous Gilbert damping from impurities and electron-electron interactions},
  author = {Hankiewicz, E. M. and Vignale, G. and Tserkovnyak, Y.},
  journal = {Phys. Rev. B},
  volume = {78},
  issue = {2},
  pages = {020404(R)},
  numpages = {4},
  year = {2008},
  month = {Jul},
  publisher = {American Physical Society},
  doi = {10.1103/PhysRevB.78.020404},
  url = {https://link.aps.org/doi/10.1103/PhysRevB.78.020404}
}

@article{Nembach2013,
  title = {Mode- and Size-Dependent Landau-Lifshitz Damping in Magnetic Nanostructures: Evidence for Nonlocal Damping},
  author = {Nembach, Hans T. and Shaw, Justin M. and Boone, Carl T. and Silva, T. J.},
  journal = {Phys. Rev. Lett.},
  volume = {110},
  issue = {11},
  pages = {117201},
  numpages = {5},
  year = {2013},
  month = {Mar},
  publisher = {American Physical Society},
  doi = {10.1103/PhysRevLett.110.117201},
  url = {https://link.aps.org/doi/10.1103/PhysRevLett.110.117201}
}

@article{Tserkovnyak2009,
  title = {Transverse spin diffusion in ferromagnets},
  author = {Tserkovnyak, Yaroslav and Hankiewicz, E. M. and Vignale, Giovanni},
  journal = {Phys. Rev. B},
  volume = {79},
  issue = {9},
  pages = {094415},
  numpages = {11},
  year = {2009},
  month = {Mar},
  publisher = {American Physical Society},
  doi = {10.1103/PhysRevB.79.094415},
  url = {https://link.aps.org/doi/10.1103/PhysRevB.79.094415}
}

@article{Hamdi23,
  title = {Spin wave dispersion of ultra-low damping hematite ($\ensuremath{\alpha}\ensuremath{-}{\mathrm{{Fe}}}_{2}{\mathrm{{O}}}_{3}$) at GHz frequencies},
  author = {Hamdi, Mohammad and Posva, Ferdinand and Grundler, Dirk},
  journal = {Phys. Rev. Mater.},
  volume = {7},
  issue = {5},
  pages = {054407},
  numpages = {7},
  year = {2023},
  month = {May},
  publisher = {American Physical Society},
  doi = {10.1103/PhysRevMaterials.7.054407},
  url = {https://link.aps.org/doi/10.1103/PhysRevMaterials.7.054407}
}

@article{Camley1982,
  title = {Stokes---anti-Stokes asymmetry in Brillouin scattering from magnons in thin ferromagnetic films},
  author = {Camley, R. E. and Gr{\"u}nberg, P. and Mayr, C. M.},
  journal = {Phys. Rev. B},
  volume = {26},
  issue = {5},
  pages = {2609--2614},
  numpages = {0},
  year = {1982},
  month = {Sep},
  publisher = {American Physical Society},
  doi = {10.1103/PhysRevB.26.2609},
  url = {https://link.aps.org/doi/10.1103/PhysRevB.26.2609}
}

@article{DamonEshbach1961,
title = {Magnetostatic modes of a ferromagnet slab},
journal = {J. Phys. Chem. Solids},
volume = {19},
number = {3},
pages = {308-320},
year = {1961},
issn = {0022-3697},
doi = {https://doi.org/10.1016/0022-3697(61)90041-5},
url = {https://www.sciencedirect.com/science/article/pii/0022369761900415},
author = {R.W. Damon and J.R. Eshbach}
}

@article{Zhang2003,
  title = {High-frequency standing longitudinal acoustic resonances in supported thin films},
  author = {Zhang, X. and Bandhu, R. S. and Sooryakumar, R. and Jonker, B. T.},
  journal = {Phys. Rev. B},
  volume = {67},
  issue = {7},
  pages = {075407},
  numpages = {5},
  year = {2003},
  month = {Feb},
  publisher = {American Physical Society},
  doi = {10.1103/PhysRevB.67.075407},
  url = {https://link.aps.org/doi/10.1103/PhysRevB.67.075407}
}

@article{KalinikosSlavin1986,
doi = {10.1088/0022-3719/19/35/014},
url = {https://doi.org/10.1088/0022-3719/19/35/014},
year = {1986},
month = {dec},
publisher = {},
volume = {19},
number = {35},
pages = {7013},
author = {B A Kalinikos and A N Slavin},
title = {Theory of dipole-exchange spin wave spectrum for ferromagnetic films with mixed exchange boundary conditions},
journal = {J. Phys. C: Solid State Phys.}
}

@article{Scarponi,
  title = {High-Performance Versatile Setup for Simultaneous Brillouin-Raman Microspectroscopy},
  author = {Scarponi, F. and Mattana, S. and Corezzi, S. and Caponi, S. and Comez, L. and Sassi, P. and Morresi, A. and Paolantoni, M. and Urbanelli, L. and Emiliani, C. and Roscini, L. and Corte, L. and Cardinali, G. and Palombo, F. and Sandercock, J. R. and Fioretto, D.},
  journal = {Phys. Rev. X},
  volume = {7},
  issue = {3},
  pages = {031015},
  numpages = {11},
  year = {2017},
  month = {Jul},
  publisher = {American Physical Society},
  doi = {10.1103/PhysRevX.7.031015},
  url = {https://link.aps.org/doi/10.1103/PhysRevX.7.031015}
}

@article{Makhfudz12,
  title = {Inertia and Chiral Edge Modes of a Skyrmion Magnetic Bubble},
  author = {Makhfudz, Imam and Kr{\"u}ger, Benjamin and Tchernyshyov, Oleg},
  journal = {Phys. Rev. Lett.},
  volume = {109},
  issue = {21},
  pages = {217201},
  numpages = {4},
  year = {2012},
  month = {Nov},
  publisher = {American Physical Society},
  doi = {10.1103/PhysRevLett.109.217201},
  url = {https://link.aps.org/doi/10.1103/PhysRevLett.109.217201}
}

@article{Chumak2015,
  author  = {Chumak, Andrii V. and Vasyuchka, Vitaliy I. and
             Serga, Alexander A. and Hillebrands, Burkard},
  title   = {Magnon spintronics},
  journal = {Nat. Phys.},
  volume  = {11},
  number  = {6},
  pages   = {453--461},
  year    = {2015},
  doi     = {10.1038/nphys3347}
}

@ARTICLE{Chumak22,
  author={Chumak, A. V. and Kabos, P. and Wu, M. and Abert, C. and Adelmann, C. and Adeyeye, A. O. and Åkerman, J. and Aliev, F. G. and Anane, A. and Awad, A. and Back, C. H. and Barman, A. and Bauer, G. E. W. and Becherer, M. and Beginin, E. N. and Bittencourt, V. A. S. V. and Blanter, Y. M. and Bortolotti, P. and Boventer, I. and Bozhko, D. A. and Bunyaev, S. A. and Carmiggelt, J. J. and Cheenikundil, R. R. and Ciubotaru, F. and Cotofana, S. and Csaba, G. and Dobrovolskiy, O. V. and Dubs, C. and Elyasi, M. and Fripp, K. G. and Fulara, H. and Golovchanskiy, I. A. and Gonzalez-Ballestero, C. and Graczyk, P. and Grundler, D. and Gruszecki, P. and Gubbiotti, G. and Guslienko, K. and Haldar, A. and Hamdioui, S. and Hertel, R. and Hillebrands, B. and Hioki, T. and Houshang, A. and Hu, C.-M. and Huebl, H. and Huth, M. and Iacocca, E. and Jungfleisch, M. B. and Kakazei, G. N. and Khitun, A. and Khymyn, R. and Kikkawa, T. and Kläui, M. and Klein, O. and Kłos, J. W. and Knauer, S. and Koraltan, S. and Kostylev, M. and Krawczyk, M. and Krivorotov, I. N. and Kruglyak, V. V. and Lachance-Quirion, D. and Ladak, S. and Lebrun, R. and Li, Y. and Lindner, M. and Macêdo, R. and Mayr, S. and Melkov, G. A. and Mieszczak, S. and Nakamura, Y. and Nembach, H. T. and Nikitin, A. A. and Nikitov, S. A. and Novosad, V. and Otálora, J. A. and Otani, Y. and Papp, A. and Pigeau, B. and Pirro, P. and Porod, W. and Porrati, F. and Qin, H. and Rana, B. and Reimann, T. and Riente, F. and Romero-Isart, O. and Ross, A. and Sadovnikov, A. V. and Safin, A. R. and Saitoh, E. and Schmidt, G. and Schultheiss, H. and Schultheiss, K. and Serga, A. A. and Sharma, S. and Shaw, J. M. and Suess, D. and Surzhenko, O. and Szulc, K. and Taniguchi, T. and Urbánek, M. and Usami, K. and Ustinov, A. B. and van der Sar, T. and van Dijken, S. and Vasyuchka, V. I. and Verba, R. and Kusminskiy, S. Viola and Wang, Q. and Weides, M. and Weiler, M. and Wintz, S. and Wolski, S. P. and Zhang, X.},
  journal={IEEE Trans. Magn.}, 
  title={Advances in Magnetics Roadmap on Spin-Wave Computing}, 
  year={2022},
  volume={58},
  number={6},
  pages={1-72},
  doi={10.1109/TMAG.2022.3149664}}

@article{bhattacharjee2012_atomistic,
  title = {Atomistic Spin Dynamic Method with Both Damping and Moment of Inertia Effects Included from First Principles},
  author = {Bhattacharjee, S. and Nordstr{\"o}m, L. and Fransson, J.},
  year = 2012,
  journal = {Phys. Rev. Lett.},
  volume = {108},
  pages = {057204},
  doi = {10.1103/PhysRevLett.108.057204}
}

@article{cherkasskii2022_theory,
  title = {Theory of inertial spin dynamics in anisotropic ferromagnets},
  author = {Cherkasskii, Mikhail and Barsukov, Igor and Mondal, Ritwik and Farle, Michael and Semisalova, Anna},
  journal = {Phys. Rev. B},
  volume = {106},
  issue = {5},
  pages = {054428},
  numpages = {10},
  year = {2022},
  month = {Aug},
  publisher = {American Physical Society},
  doi = {10.1103/PhysRevB.106.054428},
  url = {https://link.aps.org/doi/10.1103/PhysRevB.106.054428}
}

@article{cherkasskii2024_inertial,
  title = {Inertial Spin Waves in Spin Spirals},
  author = {Cherkasskii, M. and Mondal, R. and R{\'o}zsa, L.},
  year = 2024,
  journal = {Phys. Rev. B},
  volume = {109},
  pages = {184424},
  doi = {10.1103/PhysRevB.109.184424}
}

@article{ciornei2011_magnetization,
  title = {Magnetization Dynamics in the Inertial Regime: {{Nutation}} Predicted at Short Time Scales},
  author = {Ciornei, M.-C. and Rub{\'i}, J. M. and Wegrowe, J.-E.},
  year = 2011,
  journal = {Phys. Rev. B},
  volume = {83},
  pages = {020410},
  doi = {10.1103/PhysRevB.83.020410}
}

@article{de2025_magnetic,
  title = {Magnetic Nutation: {{Transient}} Separation of Magnetization from Its Angular Momentum},
  shorttitle = {Magnetic Nutation},
  author = {De, Anulekha and Schlegel, Julius and Lentfert, Akira and Scheuer, Laura and Stadtm{\"u}ller, Benjamin and Pirro, Philipp and von Freymann, Georg and Nowak, Ulrich and Aeschlimann, Martin},
  year = 2025,
  month = jan,
  journal = {Phys. Rev. B},
  volume = {111},
  number = {1},
  pages = {014432},
  issn = {2469-9950, 2469-9969},
  doi = {10.1103/PhysRevB.111.014432},
  urldate = {2026-08-28},
  langid = {english}
}

@article{fahnle2011_generalized,
  title = {Generalized {{Gilbert}} Equation Including Inertial Damping: {{Derivation}} from an Extended Breathing {{Fermi}} Surface Model},
  author = {F{\"a}hnle, M. and Steiauf, D. and Illg, C.},
  year = 2011,
  journal = {Phys. Rev. B},
  volume = {84},
  pages = {172403},
  doi = {10.1103/PhysRevB.84.172403}
}

@article{mondal2017_relativistic,
  title = {Relativistic Theory of Magnetic Inertia in Ultrafast Spin Dynamics},
  author = {Mondal, R. and Berritta, M. and Nandy, A. K. and Oppeneer, P. M.},
  year = 2017,
  journal = {Phys. Rev. B},
  volume = {96},
  pages = {024425},
  doi = {10.1103/PhysRevB.96.024425}
}

@article{neeraj2021_inertial,
  title = {Inertial Spin Dynamics in Ferromagnets},
  author = {Neeraj, K. and Awari, N. and Kovalev, S. and Polley, D. and Zhou Hagstr{\"o}m, N. and Arekapudi, S. S. P. K. and Semisalova, A. and Lenz, K. and Green, B. and Deinert, J.-C. and Ilyakov, I. and Chen, M. and Bawatna, M. and Scalera, V. and {d'Aquino}, M. and Serpico, C. and Hellwig, O. and Wegrowe, J.-E. and Gensch, M. and Bonetti, S.},
  year = 2021,
  journal = {Nat. Phys.},
  volume = {17},
  pages = {245--250},
  doi = {10.1038/s41567-020-01040-y}
}

@article{quarenta2024_bathinduced,
  title = {Bath-Induced Spin Inertia},
  author = {Quarenta, M. G. and Tharmalingam, M. and Ludwig, T. and Yuan, H. Y. and Karwacki, L. and Verstraten, R. C. and Duine, R. A.},
  year = 2024,
  journal = {Phys. Rev. Lett.},
  volume = {133},
  pages = {136701},
  doi = {10.1103/PhysRevLett.133.136701}
}

@article{unikandanunni2022_inertial,
  title = {Inertial Spin Dynamics in Epitaxial Cobalt Films},
  author = {Unikandanunni, V. and Medapalli, R. and Asa, M. and Albisetti, E. and Petti, D. and Bertacco, R. and Fullerton, E. E. and Bonetti, S.},
  year = 2022,
  journal = {Phys. Rev. Lett.},
  volume = {129},
  pages = {237201},
  doi = {10.1103/PhysRevLett.129.237201}
}

@misc{beater2026,
      title={Resonant spin mode from retarded relaxation}, 
      author={J. Beater and G. S. Uhrig},
      year={2026},
      eprint={2609.21623},
      archivePrefix={arXiv},
}

@misc{hartmann2025_intrinsic,
  title = {Intrinsic non-{Markovian} magnetisation dynamics},
  author = {Hartmann, Felix and Unikandanunni, Vivek and Bargheer, Matias and Fullerton, Eric E. and Bonetti, Stefano and Anders, Janet},
  year = {2025},
  eprint = {2512.07378},
  archiveprefix = {arXiv},
}

@article{kikuchi2015_spin,
  title = {Spin Dynamics with Inertia in Metallic Ferromagnets},
  author = {Kikuchi, T. and Tatara, G.},
  year = 2015,
  journal = {Phys. Rev. B},
  volume = {92},
  pages = {184410},
  doi = {10.1103/PhysRevB.92.184410}
}

@article{rodriguez2024_spin,
  title = {Spin Inertia and Auto-Oscillations in Ferromagnets},
  author = {Rodriguez, R. and Cherkasskii, M. and Jiang, R. and Mondal, R. and Etesamirad, A. and Tossounian, A. and Ivanov, B. A. and Barsukov, I.},
  year = 2024,
  journal = {Phys. Rev. Lett.},
  volume = {132},
  pages = {246701},
  doi = {10.1103/PhysRevLett.132.246701}
}

@misc{weissenhofer2026_spurious,
      title={Spurious spin nutation modes arising from truncated memory effects}, 
      author={Markus Weißenhofer and Ritwik Mondal and M. S. Mrudul and Peter M. Oppeneer},
      year={2026},
      eprint={2609.27487},
      archivePrefix={arXiv},
}

@article{wegrowe2012_magnetization,
    author = {Wegrowe, J.-E. and Ciornei, M.-C.},
    title = {Magnetization dynamics, gyromagnetic relation, and inertial effects},
    journal = {Am. J. Phys.},
    volume = {80},
    number = {7},
    pages = {607-611},
    year = {2012},
    month = {07},
    issn = {0002-9505},
    doi = {10.1119/1.4709188},
    url = {https://doi.org/10.1119/1.4709188},
}

@article{NikolicPRL26,
  title = {Optically Induced Magnetic Inertia and Magnons from Non-Markovian Extension of the Landau-Lifshitz-Gilbert Equation},
  author = {Reyes-Osorio, Felipe and Nikoli{\ifmmode \acute{c}\else \'{c}\fi{}}, Branislav K.},
  journal = {Phys. Rev. Lett.},
  volume = {135},
  issue = {24},
  pages = {246701},
  numpages = {10},
  year = {2025},
  month = {Dec},
  publisher = {American Physical Society},
  doi = {10.1103/sl4k-pcvq},
  url = {https://link.aps.org/doi/10.1103/sl4k-pcvq}
}

@article{BarsukovMagneticPhase,
author = {Barsukov,I.  and Fu,Yu  and Safranski,C.  and Chen,Y.-J.  and Youngblood,B.  and Gonçalves,A. M.  and Spasova,M.  and Farle,M.  and Katine,J. A.  and Kuo,C. C.  and Krivorotov,I. N. },
title = {Magnetic phase transitions in {Ta/CoFeB/MgO} multilayers},
journal = {Appl. Phys. Lett.},
volume = {106},
number = {19},
pages = {192407},
year = {2015},
doi = {10.1063/1.4921306}
}

@article{gonccalves2018oscillatory,
  title={Oscillatory interlayer coupling in spin Hall systems},
  author={Gon{\c{c}}alves, AM and Garcia, F and Lee, HK and Smith, A and Soledade, PR and Passos, CAC and Costa, M and Souza-Neto, NM and Krivorotov, IN and Sampaio, LC and others},
  journal={Sci. Rep.},
  volume={8},
  number={1},
  pages={1--6},
  year={2018},
  publisher={Nature Publishing Group}
}

@article{ Li_2016,
  title={ Wave-Number-Dependent Gilbert Damping in Metallic Ferromagnets },
  author={ Li, Y. and Bailey, W.E. },
  journal={ Phys. Rev. Lett. },
  volume={116},
  number={},
  pages={117602},
  year={2016},
  publisher={},
doi={10.1103/PhysRevLett.116.117602}
}

\end{document}